\documentclass[
  aps,
  prl,
  reprint,
  showpacs,
  groupedaddress,
  amsmath,
  amssymb,
  floatfix]{revtex4-1}

\usepackage{graphicx,epsfig,dcolumn,multirow}
\newcolumntype{d}[1]{D{.}{.}{#1}}

\RequirePackage{xspace}

\newcommand\ckmfitter{{CKMfitter}}
\newcommand{\simgt}{\,\hbox{\lower0.6ex\hbox{$\sim$}\llap{\raise0.6ex\hbox{$>$}}}\,}
\newcommand{\simlt}{\,\hbox{\lower0.6ex\hbox{$\sim$}\llap{\raise0.6ex\hbox{$<$}}}\,} 

\newcommand{\imag}{\mathrm{Im}\,}
\newcommand{\real}{\mathrm{Re}\,}
\newcommand{\dm}{\ensuremath{\Delta M}}
\newcommand{\dg}{\ensuremath{\Delta \Gamma}}
\newcommand{\ov}[1]{\overline{#1}}

\newcommand{\Bag}{\mathcal{B}}

\newcommand{\fig}[1]{Fig.~\ref{#1}} 
\newcommand{\bb}{\ensuremath{B\!-\!\Bbar{}\,}}

\newcommand{\kk}{\ensuremath{K\!-\!\Kbar{}\,}}
\newcommand{\bbms}{\bbs\ mixing}
\newcommand{\bbmd}{\bbd\ mixing}

\newcommand{\bbm}{\bb\ mixing}

\newcommand{\kkm}{\kk\ mixing}
\newcommand{\bbd}{\ensuremath{B_d\!-\!\Bbar{}_d\,}}
\newcommand{\bbs}{\ensuremath{B_s\!-\!\Bbar{}_s\,}}
\newcommand{\bbq}{\ensuremath{B_q\!-\!\Bbar{}_q\,}}
\newcommand{\Bbar}{\,\overline{\!B}}

\newcommand{\Kbar}{\,\overline{\!K}}

\newcommand{\etal}{\emph{et al.}\xspace}

\newcommand{\arxiv}[1]{{arxiv:{#1}}}

\newcommand{\nn}{\nonumber\\}

\usepackage{pstricks}
\newcmykcolor{darkgreen}{1 0 0.6 0.5}  

\def\journalL#1#2#3#4#5{\journal{#1 #2}{#3}{#4}{#5}}
\def\journal#1#2#3#4{#1~{\bf #2}, #3 (#4)}
\begin{document}

\title{
   New Physics in
    $\boldsymbol{B}$--$\boldsymbol{\overline{B}}$ mixing in the light of recent LHCb data}

\author{A.~Lenz$^{\,a}$, U.~Nierste$^{\,b}$ and\\
\vspace{0.1cm}
J.~Charles$^{\,c}$,
S.~Descotes-Genon$^{\,d}$,
H.~Lacker$^{\,e}$,
S.~Monteil$^{\,f}$,
V.~Niess$^{\,f}$,
S.~T'Jampens$^{\,g}$
[for the \ckmfitter\  Group]\\
\vspace{0.6cm}}

\affiliation{
\mbox{$^{a}$ CERN - Theory Division, PH-TH, Case C01600, CH-1211 Geneva 23,
                {e-mail: alenz@cern.ch}}\\
\mbox{$^{b}$ Institut f\"ur Theoretische Teilchenphysik,
Karlsruhe Institute of Technology, D-76128 Karlsruhe, Germany,}
\mbox{email: nierste@particle.uni-karlsruhe.de}\\   
\mbox{$^{c}$ Centre de Physique Th\'eorique,
Aix-Marseille Univ., CNRS UMR 7332, Univ.
Sud Toulon Var,}              
                \mbox{{F-13288 Marseille cedex 9, France, e-mail: charles@cpt.univ-mrs.fr}} \\
\mbox{$^{d}$ Laboratoire de Physique Th\'eorique,
 B\^{a}timent 210, Universit\'e  Paris-Sud 11, F-91405 Orsay Cedex, France} \\
\mbox{(UMR 8627 du CNRS  associ\'ee \`a l'Universit\'e Paris-Sud 11),
 {e-mail: Sebastien.Descotes-Genon@th.u-psud.fr}}\\
\mbox{$^{e}$ Humboldt-Universit\"at zu Berlin,
                   Institut f\"ur Physik,
                   Newtonstr. 15,
                   D-12489 Berlin, Germany,}
\mbox{e-mail: lacker@physik.hu-berlin.de}\\
\mbox{$^{f}$ Laboratoire de Physique Corpusculaire de Clermont-Ferrand,
                  Universit\'e Blaise Pascal,}
\mbox{24 Avenue des Landais F-63177 Aubiere Cedex,
(UMR 6533 du CNRS-IN2P3 associ\'ee \`a
                   l'Universit\'e Blaise Pascal),}
\mbox{e-mail: monteil@in2p3.fr, niess@in2p3.fr} \\
\mbox{$^{g}$ Laboratoire d'Annecy-Le-Vieux de Physique des Particules,
                      9 Chemin de Bellevue, BP 110,} 
\mbox{F-74941 Annecy-le-Vieux Cedex, France,}\\
\mbox{(UMR 5814 du CNRS-IN2P3 associ\'ee \`a
                   l'Universit\'e de Savoie),
                {e-mail: tjamp@lapp.in2p3.fr}} \\
}  
\date{\today}

\begin{abstract}
We perform model-independent statistical analyses
of three scenarios accommodating New Physics (NP) in $\Delta F=2$
flavour-changing neutral current amplitudes.  In a scenario in
  which NP in \bbmd\ and \bbms\ is uncorrelated, we find the parameter
  point representing the Standard-Model disfavoured by  2.4 standard
  deviations. However, recent LHCb data on $B_s$ neutral-meson mixing
  forbid a good accommodation of the D\O\ data on the semileptonic CP
  asymmetry $A_{\rm SL}$. We introduce a fourth scenario with NP in
both $M_{12}^{d,s}$ and $\Gamma_{12}^{d,s}$, which can accommodate
  all data. We discuss the viability  of this possibility and
  emphasise the importance of separate measurements of the CP
  asymmetries in semileptonic $B_d$ and $B_s$ decays.  All results have
been obtained with the \ckmfitter\ analysis package, featuring the
frequentist statistical approach and using Rfit to handle theoretical
uncertainties.
\end{abstract}


\maketitle

Flavour physics looks back to a quarter-century of precision studies at
the B-factories 
with a parallel
theoretical effort addressing the Standard Model (SM) predictions
for the measured quantities \cite{Buras:2011we}.  With the parameters of
the Cabibbo-Kobayashi-Maskawa (CKM) matrix \cite{Cabibbo:1963yz}
overconstrained by many measurements one can  predict
yet unmeasured quantities
\cite{Charles:2011va}.  
 Still, the global fit to the CKM unitarity triangle reveals some
discrepancies with the SM, driven by a conflict between $B(B\to
  \tau \nu)$ and $\sin(2\beta)$ measured from $B_d \to J/\Psi K$
  \cite{Lenz:2010gu,Lunghi:2010gvBona:2009cj}. 
Furthermore, in May 2010 the D\O\ experiment reported a deviation
  of the semileptonic CP asymmetry (dimuon asymmetry) in $B_{d,s}$
  decays from its SM prediction \cite{bbln,ln} by 3.2$\,\sigma$
  \cite{dimuon_evidence_d0}.  In June 2011 this discrepancy has
  increased to 3.9$\,\sigma$ \cite{Abazov:2011yk}. In summer 2010 the
  data could be interpreted in well-motivated scenarios with New Physics
  (NP) in \bbm\ amplitudes \cite{Lenz:2010gu}. In this letter we present novel analyses
  which include the new data of 2011, in particular from the LHCb
  experiment.

\bbq $(q=d,s)$ oscillations involve the off-diagonal elements 
$M_{12}^q$ and $\Gamma_{12}^q$ of the
$2\times 2$ mass and decay matrices, respectively.
One can fix the three physical quantities $|M_{12}^q|$,
  $|\Gamma_{12}^q|$ and $\phi_q=\arg(-M_{12}^q/\Gamma_{12}^q)$ from the
mass difference $\dm_q\simeq 2|M_{12}^q| $ among the eigenstates, their
width difference $\dg_q \simeq 2\, |\Gamma_{12}^q| \cos \phi_q$ and the
semileptonic CP asymmetry
\begin{eqnarray}
a^q_{\rm SL} &=& \imag \frac{\Gamma_{12}^q}{M_{12}^q} =
\frac{|\Gamma_{12}^q|}{|M_{12}^q|} \sin \phi_q \; = \;
\frac{\dg_q}{\dm_q} \tan \phi_q . \label{defafs}
\end{eqnarray}
$M_{12}^q$ is especially sensitive to NP.
Therefore the two complex parameters 
$\Delta_s$ and 
$\Delta_d$, defined as   
\begin{eqnarray}
M_{12}^q & \!\equiv\! & M_{12}^{\text{SM},q} \cdot  \Delta_q \, ,
\quad\;  \Delta_q  \equiv   |\Delta_q| e^{i \phi^\Delta_q} , 
\quad\; q=d,s, \; \label{defdel}
\end{eqnarray}
can differ substantially from the SM value $\Delta_s=\Delta_d=1$. 
Importantly, the NP phases $\phi^\Delta_{d,s}$ do not only
affect $a^{d,s}_{\rm SL}$, but also shift the CP phases extracted from
the mixing-induced CP asymmetries in $B_d \to J/\Psi K$ and $B_s \to
J/\Psi \phi$ to $2\beta+\phi^\Delta_d$ and $2\beta_s-\phi^\Delta_s$,
respectively.  In summer 2010 the CDF and D\O\ analyses of $B_s \to
J/\Psi \phi$ pointed towards a large negative value of $\phi^\Delta_s$,
while simultaneously being consistent with the SM due to large
errors. With a large $\phi^\Delta_s<0$ we could  
  accommodate D\O's large negative value for the semileptonic CP
  asymmetry reading $A_{\rm SL} = 0.6 a_{\rm SL}^d + 0.4 a_{\rm SL}^s$
  in terms of the individual semileptonic CP asymmetries in the $B_d$
  and $B_s$ systems.
Moreover, the discrepancy between $B(B\to \tau \nu)$
and the mixing-induced CP asymmetry in $B_d \to J/\Psi K$ 
can be removed with
$\phi_d^\Delta<0$.  The allowed range for $\phi_d^\Delta$ implies a
  contribution to $A_{\rm SL}$ with the right (i.e.\ negative)
sign.  In our 2010 analysis in Ref.~\cite{Lenz:2010gu} we have
determined the preferred ranges for $\Delta_s$ and $\Delta_d$ in a
simultaneous fit to the CKM parameters in three generic scenarios in
which NP is confined to $\Delta F=2$ flavour-changing neutral currents. 
In our Scenario~I we have treated  $\Delta_s$, $\Delta_d$ (and three
  more parameters related to \kkm ) independently, corresponding to
NP with arbitrary flavour structure. Scenario II implements
minimal-flavour violation (MFV) with small bottom Yukawa coupling
entailing real $\Delta_s=\Delta_d$. Scenario III covers MFV models in
which $\Delta_s=\Delta_d$ is allowed to be complex. In
Ref.~\cite{Lenz:2010gu} we have found an excellent fit in Sc.~I (and a
good fit in Sc.~3) with all discrepancies relieved through
$\Delta_{d,s}\neq 1$, while the fit has returned \kkm\ essentially
SM-like.
 
The recent LHCb measurement of  
  the CP phase $\phi_s^{\psi\phi}$ from $A_{\rm CP}^{\rm mix} (B_s \to
  J/\Psi \phi)$ does not permit large deviations of $\phi_s^\Delta$ from
  zero anymore.
This trend was
  also confirmed by the latest CDF results~\cite{CDF:2011af}. 
The current situation with the phase $2\phi_s^{\psi\phi}
\equiv -2\beta_s + \phi_s^\Delta$ and $A_{\rm SL}$ is as follows (at 68\% CL):
\begin{align}
   2 \phi_s^{\psi\phi} = (-32^{+22}_{-21})^\circ
  \quad & \mbox{D\O\ \cite{taggedphaseD0_2}} \nn 
-60^\circ \leq  2  \phi_s^{\psi\phi}  \leq -2.3^\circ 
  \quad & \mbox{CDF \cite{CDF:2011af}} \nn  
  2  \phi_s^{\psi\phi} = (-0.1\pm 5.8\pm 1.5)^\circ 
   \quad & \mbox{LHCb}  \mbox{ $J/\psi \phi$ \cite{LHCb:2011aa}}  \nn
  2  \phi_s^{\psi f_0} = (-25.2\pm 25.2\pm 1.2)^\circ
   \quad & \mbox{LHCb}  \mbox{ $J/\psi f_0$ \cite{LHCb:2011ab}} \nn
A_{\rm SL} = 
 (-7.87 \pm 1.72  \pm 0.93 ) \cdot 10^{-3} 
   \quad & \mbox{D\O\ \cite{Abazov:2011yk}} 
 \label{exnum}  
\end{align}
Here $2\beta_s=2\arg (-V_{ts} V_{tb}^*/(V_{cs} V_{cb}^*))\simeq 2.2^\circ$~\cite{CKMfitterwebsite}.

From this discussion, there is a conflict between LHCb data on $B_s\to J/\psi
\phi$ and the D\O\ measurement of $A_{\rm SL}$ which we cannot fully
resolve in our Scenarios I, II and III.  We therefore discuss a fourth scenario which
also permits NP in the decay matrices $\Gamma_{12}^s$ or
$\Gamma_{12}^d$.

\section{Results for Scenarios~I, II and III}
In Tab.~\ref{tab:Inputs} we summarise the changes in the inputs
compared to Tabs.~1--7 of Ref.~\cite{Lenz:2010gu}.  Following
Ref.~\cite{Charles:2011va} we have included $K_{\ell 3}$, $K_{\ell 2}$,
$\pi_{\ell 2}$ (and the related $\tau$ decays) for $|V_{ud}|$ and $|V_{us}|$.  Concerning the measurements of
$(\phi_s,\Gamma_s)$ from $B_s\to J/\psi \phi$, we have combined
  the CDF and LHCb results by taking the product of their 2D
  profile-likelihoods~\cite{LHCb:2011aa,CDF:2011af}.  Unfortunately, we
could not obtain the corresponding likelihood from {D\O}.
The impact of this omission is mild due to the smaller
uncertainties of the CDF and LHCb results. 
We have neither used the LHCb result on $B_s\to J/\psi f_0$
as only $\phi_s$ (not the 2D likelihood) was provided in Ref.~\cite{LHCb:2011ab}.  But we have
included the flavour-specific $B_s$
  lifetime $\tau^{FS}_{B_s}$~\cite{HFAG11} providing an
  independent constraint on $\Delta\Gamma_s$.  
We analyse the
{D\O} measurement of $A_{\rm SL}$ with the
production fractions at 1.8-2 TeV according to Ref.~\cite{HFAG11}: $f_s=
0.111\pm 0.014$ and $f_d=0.339\pm 0.031$, 
corresponding to
$A_{\rm SL}=(0.532\pm 0.039) a^{d}_{\rm SL}+ (0.468\pm 0.039) a^{s}_{\rm
    SL}$.

\begin{table}[Htp]
\renewcommand{\arraystretch}{1.3}
\centering
\begin{tabular}{|c|c|c|}\hline
Observable                                      & Value and uncertainties                      & Ref.\              \\
\hline
$\mathcal{B}(K \rightarrow e \nu_{e})$    & $(1.584 \pm 0.020) \times 10^{-5}$  & \cite{PDG}\\
$\mathcal{B}(K \rightarrow \mu \nu_{\mu})$    & $0.6347\pm 0.0018$  & \cite{Antonelli:2010yf}\\
$\mathcal{B}(\tau \rightarrow K \nu_{\tau})$    & $0.00696\pm   0.00023$  & \cite{Antonelli:2010yf}\\
$\mathcal{B}(K \!\rightarrow \mu \nu_{\mu})/\mathcal{B}(K \!\rightarrow \pi \nu_{\mu})$    & $1.3344\pm 0.0041$  & \cite{Antonelli:2010yf}\\
$\mathcal{B}(\tau \rightarrow K \nu_{\tau})/\mathcal{B}(\tau \rightarrow \pi \nu_{\tau})$    & $(6.53\pm 0.11)\cdot 10^{-2}$  & \cite{Banerjee:2008hg}\\
$\alpha$                                        &  $88.7{^{\,+4.6}_{\,-4.3}}^{\circ}$          & \cite{CKMfitterwebsite}              \\
$\gamma$                                        & $(66\pm 12)^{\circ}$              & \cite{CKMfitterwebsite}              \\
\hline
$\Delta m_{d}$                                  & $0.507  \pm 0.004 {\rm ps}^{-1}$                   &\cite{PDG}          \\
$\Delta m_{s}$                                  & $17.731\pm 0.045 {\rm ps}^{-1}$                    &\cite{Aaij:2011qx,Abulencia:2006ze}      \\
$A_\text{SL}$                                   & $(-74 \pm 19)\times 10^{-4}$                       & \cite{Abazov:2011yk} \\
$\phi_s^{\psi\phi}$ vs. $\Delta \Gamma_{s}$ & see text                                     & \cite{LHCb:2011aa,CDF:2011af}
                   \\ 
\hline
\end{tabular}

$\quad$

\begin{tabular}{|c|c|c|}\hline
  Theoretical Parameter                          & Value and uncertainties                 & Ref.\                                 \\
  \hline
  $f_{B_{s}}$ &  $229\pm2 \pm6$ MeV &  \cite{CKMfitterwebsite}\\
  $f_{B_{s}}/f_{B_{d}}$                          & $1.218   \pm 0.008 \pm 0.033$          & \cite{CKMfitterwebsite}  \\
  $\widehat\Bag_{B_{s}}$                  & $1.291  \pm 0.025 \pm 0.035$           &  \cite{CKMfitterwebsite}\\
  $\Bag_{B_{s}}/\Bag_{B_{d}}$                    & $1.024    \pm 0.013  \pm 0.015$            &   \cite{CKMfitterwebsite}\\
  ${\hat{\Bag}}_{K}$                   &  $(0.733  \pm 0.003 \pm 0.036)$
        &   \cite{CKMfitterwebsite}\\
  $f_{K}$       &  $156.3 \pm 0.3 \pm 1.9  {\rm MeV}$    
                &       \cite{CKMfitterwebsite} \\ 
  $f_{K}/f_\pi$                               & $1.1985\pm 0.0013\pm 0.0095$ &  \cite{CKMfitterwebsite}\\
  $\alpha_s(M_{Z})$             & $ 0.1184\pm 0 \pm 0.0007$                 & \cite{PDG}  \\
  \hline
\end{tabular}
\caption
{Experimental and theoretical inputs inputs added or modified 
compared to Ref.~\cite{Lenz:2010gu} and used in our fits.}
\label{tab:Inputs}
\end{table}

\begin{figure}
  \includegraphics[width=8cm]{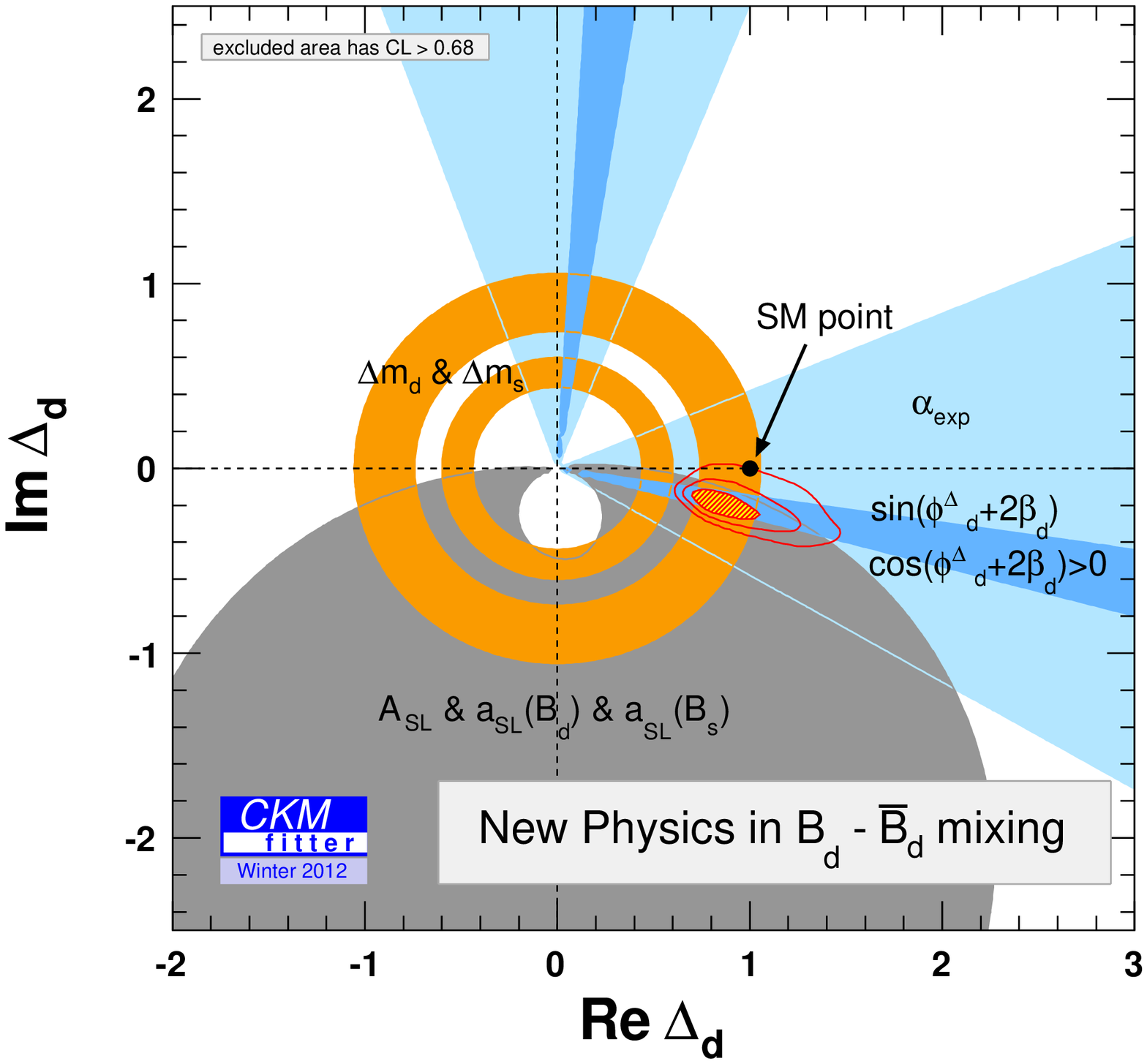}
   \includegraphics[width=8cm]{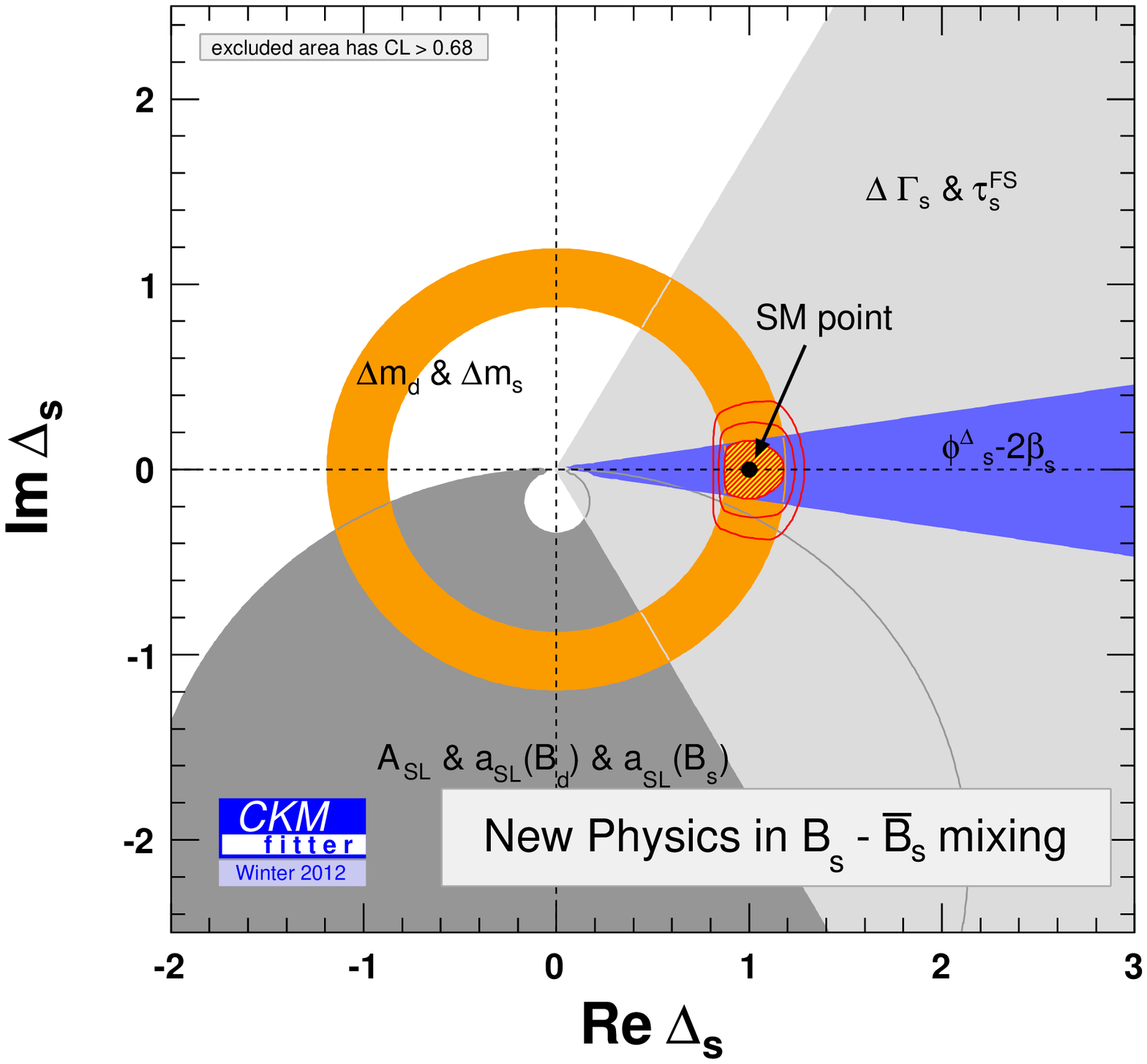}
\caption{\small Complex parameters  $\Delta_d$  (up) 
     and $\Delta_s$ (down) in Scenario I.  
     Here $\alpha_{\rm exp}\equiv \alpha -\phi_d^\Delta/2$.
     The coloured areas represent regions
     with ${\rm CL} < 68.3~\%$ for the individual constraints. 
     The red area shows
     the region with ${\rm CL} < 68.3~\%$ for the combined fit, 
    with the two additional contours 
     delimiting the regions with ${\rm CL} < 95.45~\%$ and 
     ${\rm CL} < 99.73~\%$.   The $p$-value for the 2D SM hypothesis $\Delta_d=1$ ($\Delta_s=1$) is
     3.0 $\sigma$ (0.0 $\sigma$).
     \label{fig-Delta_scenario1}}
\end{figure}

\begin{figure}
\includegraphics[width=8cm]{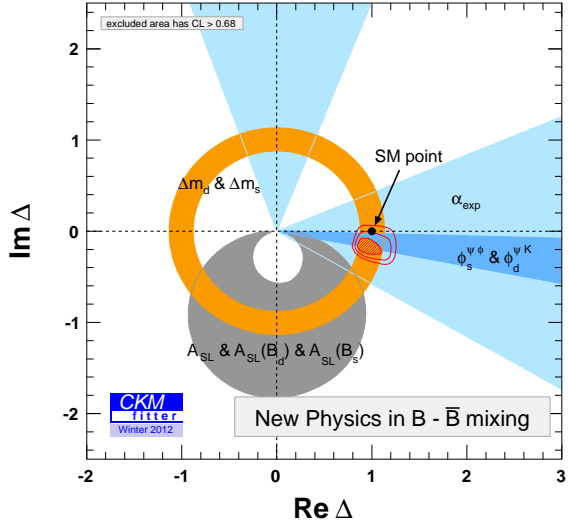}
 \caption{\small Constraint on the complex parameter $\Delta \equiv \Delta_d=\Delta_s$ from 
                  the fit in Scenario III with same conventions as in fig.~\ref{fig-Delta_scenario1}.
                                    The $p$-value for the 2D SM hypothesis $\Delta=1$ is 2.1 $\sigma$. \label{fig-Delta_scenario3}}
\end{figure}

We summarise our results  in
Tabs.~\ref{tab-results} and \ref{tab-pulls} and in
\fig{fig-Delta_scenario1} 
(Sc.~I) as well as
\fig{fig-Delta_scenario3} (Sc.~III).
Even in Sc.~I  our fit to the
data  is significantly worse than in 2010~\cite{Lenz:2010gu}:
While $\phi_d^\Delta<0 $ alleviates the discrepancy of $A_{\rm SL}$ with
the SM, the LHCb result on $\phi_s^{\psi\phi}$ prevents larger
contributions from the $B_s$ system to $A_{\rm SL}$. In Sc.~I, we find
pull values for $A_{\rm SL}$ and $\phi_s^\Delta-2\beta_s$ of
 3.0$\,\sigma$ and 2.7$\,\sigma$ respectively (compared to
1.2$\,\sigma$ and 0.5$\,\sigma$ in Ref.~\cite{Lenz:2010gu}).  We do not
quote pull values for $\Delta m_{d,s}$ in Sc.~I, as these observables
are not constrained
once their experimental measurement is removed.  In contrast to earlier analyses,
only one solution for  $\Delta_s$ survives
thanks to 
the recent LHCb determination of 
$\Delta\Gamma_s>0$~\cite{LHCbDGsSign} entailing $\real
  \Delta_s>0$.
Tab.~\ref{tab:pvalues} lists the p-values for various SM
  hypotheses within our NP Scenarios (more information
can be found in
Ref.~\cite{CKMfitterwebsite}).

\begin{table}
\begin{tabular}{lcc} \hline
Quantity   & $1 \sigma$ & $3\sigma$ \\
 \hline  &&   \\[-0.3cm]
$\mbox{Re}{(\Delta_d)}$  &  $0.823^{+0.143}_{-0.095}$ & $0.82^{+0.54}_{-0.20}$ \\[0.15cm]
$\mbox{Im}{(\Delta_d)}$  &  $-0.199^{+0.062}_{-0.048}$ & $-0.20^{+0.18}_{-0.19}$ \\[0.15cm]
$|\Delta_d|$             &  $0.86^{+0.14}_{-0.11}$ & $0.86^{+0.55}_{-0.22}$ \\[0.15cm]
$\phi^\Delta_d$ [deg]    & $-13.4^{+3.3}_{-2.0}$ & $-13.4^{+12.1}_{-6.0}$ \\[0.15cm]
$\mbox{Re}{(\Delta_s)}$  &  $0.965^{+0.133}_{-0.078}$ & $0.97^{+0.30}_{-0.13}$ \\[0.15cm]
$\mbox{Im}{(\Delta_s)}$  &  $-0.00^{+0.10}_{-0.10}$ & $-0.00^{+0.32}_{-0.32}$ \\[0.15cm]
$|\Delta_s|$             &  $0.977^{+0.121}_{-0.090}$ & $0.98^{+0.29}_{-0.15}$ \\[0.15cm]
$\phi^\Delta_s$ [deg]    & $-0.1^{+6.1}_{-6.1}$ & $-0^{+18.}_{-18.}$ \\[0.15cm]
$\phi^\Delta_d+2\beta$ [deg] (!)   &  $17^{+12.}_{-13.}$ & $17^{+40.}_{-55.}$ \\[0.15cm]
$\phi^\Delta_s-2\beta_s$ [deg] (!) &  $-56.8^{+10.9}_{-7.0}$ & $-57.^{+66.}_{-20.}$ \\[0.15cm]
\hline &&      \\[-0.3cm]
$A_\text{SL}$~~$[10^{-4}]$ (!)    & $-15.6^{+9.2}_{-3.9}$ & $-16^{+19}_{-12}$ \\[0.15cm]
$A_\text{SL}$~~$[10^{-4}]$        & $-17.7^{+3.9}_{-3.8}$ & $-18^{+15}_{-12}$ \\[0.15cm]
$a_\text{SL}^{s}-a_\text{SL}^{d}$~~$[10^{-4}]$             
                & $33.6^{+7.5}_{-8.2}$ & $34^{+24}_{-32}$ \\[0.15cm]
$a_\text{SL}^{d}$~~$[10^{-4}]$ (!)            
                & $-33.2^{+6.6}_{-4.1}$ & $-33^{+25}_{-13}$ \\[0.15cm]
$a_\text{SL}^{s}$~~$[10^{-4}]$ (!)            
                & $0.4^{+6.2}_{-6.3}$ & $0^{+20}_{-21}$ \\[0.15cm]
$\Delta\Gamma_d [\mathrm{ps}^{-1}]$             
                & $0.00480^{+0.00070}_{-0.00129}$ & $0.0048^{+0.0020}_{-0.0031}$ \\[0.15cm]
$\Delta\Gamma_s [\mathrm{ps}^{-1}]$ (!)            
                & $0.155^{+0.020}_{-0.079}$ & $0.155^{+0.036}_{-0.098}$ \\[0.15cm]
$\Delta\Gamma_s [\mathrm{ps}^{-1}]$            
                & $0.104^{+0.017}_{-0.016}$ & $0.104^{+0.052}_{-0.041}$ \\[0.15cm]
\hline &&      \\[-0.3cm]
$B\to \tau\nu$~~$[10^{-4}]$ (!) & $1.341^{+0.064}_{-0.232}$ & $1.34^{+0.20}_{-0.73}$ \\[0.15cm]
$B\to \tau\nu$~~$[10^{-4}]$     & $1.354^{+0.063}_{-0.095}$ & $1.35^{+0.19}_{-0.50}$ \\[0.15cm]
\hline
\end{tabular}

\caption{CL intervals for the results of the fits in Scenario I. The notation (!) means that the fit output represents the indirect constraint with the corresponding direct input  removed.\label{tab-results}}
\end{table}

\begin{table}
\begin{center}
\begin{tabular}{lcccc} \hline
Quantity & & Deviation  & wrt& \\ 
& SM & Sc. I & Sc. II &  Sc. III \\
 \hline \\[-0.3cm]
$\phi^\Delta_d+2\beta$ & $2.7~\sigma$ & $2.1~\sigma$& $2.7~\sigma$ & $1.2~\sigma$\\[0.15cm]
$\phi^\Delta_s-2\beta_s$ & $0.3~\sigma$ & $2.7~\sigma$& $0.3~\sigma$ & $2.4~\sigma$\\[0.15cm]
\hline &      \\[-0.3cm]
$|\epsilon_K|$ & $0.0~\sigma$ & - & $0.0~\sigma$ & - \\[0.15cm]
$\Delta m_d$ & $1.0~\sigma$ &  - 
                & $1.0~\sigma$ & $0.9~\sigma$ \\[0.15cm]
$\Delta m_s$ & $0.0~\sigma$ & -
                & $1.0~\sigma$ & $1.3~\sigma$ \\[0.15cm]
$A_\text{SL}$ & $3.7~\sigma$ & $3.0~\sigma$ & $3.7~\sigma$ & $3.0~\sigma$ \\[0.15cm]
$a_\text{SL}^{d}$ & $0.9~\sigma$ & $0.3~\sigma$ & $0.8~\sigma$ & $0.4~\sigma$ \\[0.15cm]
$a_\text{SL}^{s}$ & $0.2~\sigma$ & $0.2~\sigma$ & $0.2~\sigma$ & $0.0~\sigma$ \\[0.15cm]
$\Delta\Gamma_s$ & $0.0~\sigma$ & $0.4~\sigma$ & $0.0~\sigma$ & $1.0~\sigma$ \\[0.15cm]
\hline &&&& \\[-0.3cm]
$\mathcal{B}(B\to\tau\nu)$ & $2.8~\sigma$ & $1.1~\sigma$ & $2.8~\sigma$ & $1.7~\sigma$ \\[0.15cm]
\hline &&&& \\[-0.3cm]
$\mathcal{B}(B\to\tau\nu)$, $A_\text{SL}$ & $4.3~\sigma$ & $2.8~\sigma$ & $4.2~\sigma$ & $3.4~\sigma$ \\[0.15cm]
$\phi_s^\Delta-2\beta_s$, $A_\text{SL}$ & $3.3~\sigma$ & $2.7~\sigma$ & $3.3~\sigma$ & $3.2~\sigma$ \\[0.15cm]
$\mathcal{B}(B\to\tau\nu)$, $\phi_s^\Delta-2\beta_s$, $A_\text{SL}$ & $4.0~\sigma$ & $2.4~\sigma$ & $3.9~\sigma$ & $3.2~\sigma$ \\[0.15cm]
\hline
 \end{tabular} 
\end{center}
\caption{Pull values for selected parameters and observables in SM and Scenarios I, II, III, in terms of the number of 
equivalent standard deviations between the direct measurement and the full indirect
fit predictions.}\label{tab-pulls}
\end{table}  

\begin{table}
\begin{center}
\begin{tabular}{lccc}
\hline
Hypothesis & Sc.~I & Sc.~II & Sc.~III \\
\hline
$\mathrm{Im}{\Delta_d}=0$  & $3.2 \sigma$ & & $2.6 \sigma$\\
$\mathrm{Im}{\Delta_s}=0$  & $0.0 \sigma$ & & \\
\hline
$\Delta_d=1$  & $3.0 \sigma$ & $0.6 \sigma$ & $2.1 \sigma$\\
$\Delta_s=1$  & $0.0 \sigma$ & & \\
$\mathrm{Im}{\Delta_d}=\mathrm{Im}{\Delta_s}=0$  & $2.8 \sigma$ & & \\
\hline
$\Delta_d=\Delta_s=1$ &  $2.4 \sigma$ & & \\
\hline
\end{tabular}
\end{center}
\caption{p-values for various Standard Model hypotheses in the framework
  of three  NP Scenarios considered. These numbers are computed from the
  $\chi^2$ difference with and without the hypothesis constraint,
  interpreted with the appropriate number of degrees of
  freedom.\label{tab:pvalues}}
\end{table}
 
 \begin{figure}
  \includegraphics[width=8cm]{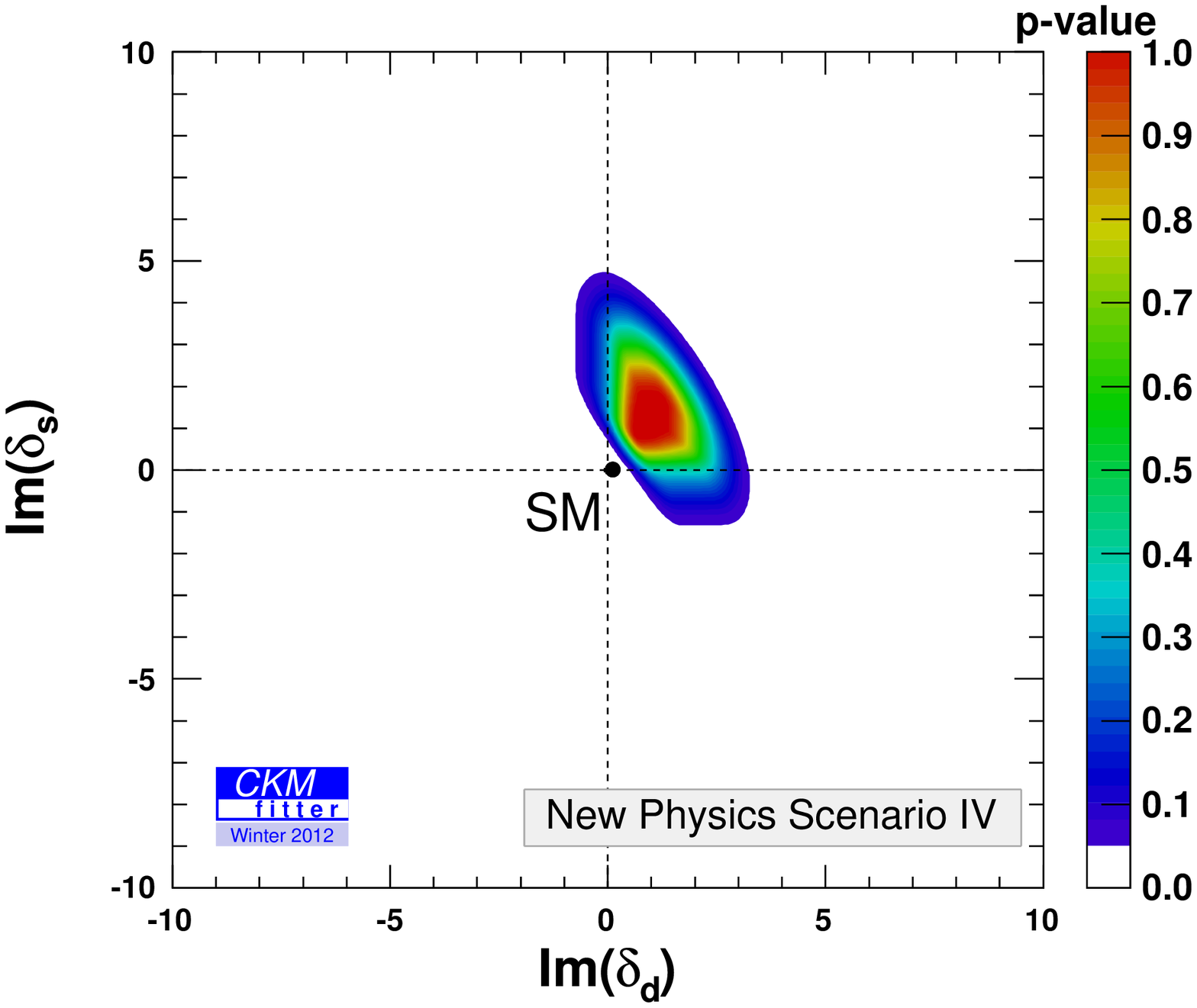}
  \caption{\small Constraints on $\imag\delta_d,\imag\delta_s$ in
    Scenario IV. The 1D 68\%CL intervals are  $\imag\delta_d=0.92^{+1.13}_{-0.69},\ \imag\delta_s=1.2^{+1.6}_{-1.0}$. 
    The $p$-value for the 2D SM hypothesis $\imag\delta_d=0.097,\imag\delta_s=-0.0057$  is
    3.2 $\sigma$.
                  \label{fig-ScenarioIV}}
                  \end{figure}

\boldmath
\section{New Physics in $\Gamma_{12}^s$ or $\Gamma_{12}^d$}
\unboldmath Several authors have discussed the possibility of a sizable
new CP-violating contribution to $\Gamma_{12}^s$ to explain the D\O\
measurement of $A_{SL}$ \cite{nping12} by
postulating new $B_s$ decay channels with large branching fraction. In such models also the
width difference $\dg_s$ typically deviates from the SM prediction in
Ref.~\cite{dega,ln,Lenz:2011ti}. $\Gamma_{12}^s$ is dominated by the
CKM-favoured tree-level decay $b\to c\bar{c}s$. Any competitive new
decay mode will increase the total $B_s$ width, which LHCb finds as
$\Gamma_s= 0.657 \pm 0.009 \pm 0.008 $ \cite{LHCb:2011aa}, implying
$\Gamma_s/\Gamma_d= 0.998 \pm 0.014 \pm 0.012 $ in excellent agreement
with the SM expectation $0\leq \Gamma_s/\Gamma_d-1\leq 4\cdot 10^{-4}$
\cite{Lenz:2011ti}. The new interaction will open new $b\to s$
decay modes affecting precisely measured inclusive $B_d$ and $B^+$ quantities \cite{Lenz:2010gu}. Furthermore, new decays mediated
by a particle with mass $M>M_W$ will add a term of order $M_W^4/M^4$
to $\Gamma_{12}^s/\Gamma_{12}^{{\rm SM},s}$, while $\Delta_s$ normally
receives a larger contribution of order $M_W^2/M^2$. In models involving
a fermion pair $(f,\ov f)$ in the final state, e.g.\ those with an
enhanced $B_s \to \tau \ov \tau$ decay \cite{nping12}, one can solve
this problem through chirality suppression. The extra contribution to
$M_{12}^s$ is down by another factor of $m_f^2/M^2$, while that to
$\Gamma_{12}^s$ is affected by the milder factor of $m_f^2/m_b^2$.
Quantities like $\Gamma_{d,s}$ will not be
chirality suppressed.  Therefore it seems not possible to add large NP effects to $\Gamma_{12}^s$.

 Phenomenologically it is thus much easier to postulate
NP in $\Gamma_{12}^d$ rather than $\Gamma_{12}^s$, because
$\Gamma_{12}^d$ is constituted by Cabibbo-suppressed decay modes like
$b\to c \ov c d$. Also here chirality suppression is welcome to
avoid problems with $M_{12}^d$, but inclusive decay observables like the
semileptonic branching fraction or the unmeasured $\dg_d$ pose no
danger. Clearly, testing this hypothesis calls for a better measurement
of $a_{\rm SL}^d$. We have studied a Scenario IV including the
possibility of NP in $\Gamma_{12}^{d,s}$.
  We stress that Sc.~IV permits NP in the $|\Delta F|=1$
  transitions contributing to $\Gamma_{12}^q$, but not in other $|\Delta
  F|=1$ quantities entering our fits, such as ${\cal B} (B\to \tau
  \nu)$. Further no new CP phase in $b\to c \ov{c} s$, which would
  change $\phi_{d,s}^\Delta$, is considered. Such a phase might further
  increase the hadronic uncertainty from penguin pollution, which is not
  an issue in the SM at the current levels of experimental precision.
 
Handy new parameters are 
\begin{align}
\!\delta_q = \frac{\Gamma_{12}^q/M_{12}^q}{\real
             (\Gamma_{12}^{{\rm SM},q}/M_{12}^{{\rm SM},q})},  \quad q=d,s,
\end{align}
$\real\delta_q$, $\imag\delta_q$ amount to
$(\dg_q/\dm_q)/(\dg_q^{\rm SM}/\dm_q^{\rm SM})$ and $-a_{\rm
  SL}^q/(\dg_q^{\rm SM}/\dm_q^{\rm SM})$, respectively. The best fit
  values of the SM predictions are  $\delta_d^{\rm SM}=1 + 0.097\, i$ and $ \delta_s^{\rm SM}= 1 - 0.0057\, i$. $\real\delta_d$ is experimentally
only weakly constrained. We illustrate  the correlation between 
$\imag \delta_d $ and $\imag \delta_s$ in \fig{fig-ScenarioIV}, 
relegating correlations of $\real \delta_s$ with $\imag \delta_{d,s}$
to  Ref.~\cite{CKMfitterwebsite}).  
The p-value of the 8D SM hypothesis $\Delta_d=\Delta_s=1$,  $\delta_{d,s}=\delta_{d,s}^{\rm SM}$
is  2.6 $\sigma$.

We stress that too large values
for $|\delta_s-\delta_s^{\rm SM}|$ are in conflict with other
observables as explained above. 
We have also studied Scenario IV without NP in the $B_s$ sector ($\Delta_s=1$ and $\delta_s=\delta_{s,\mathrm{SM}}$). 
It could accommodate the main anomalies by improving the fit by  $3.3\sigma$, but with large contributions to $\Gamma^d_{12}$: $\imag\delta_d=1.60^{+1.02}_{-0.76}$.

\section{Conclusions}
We have performed new global fits to flavour physics data 
in scenarios with generic NP in the \bbd\ and \bbms\ amplitudes,
as defined in Ref.~\cite{Lenz:2010gu}. Our results represent the status
of the end of the year 2011. Unlike in summer 2010 the two complex NP
parameters $\Delta_d$ and $\Delta_s$ (parametrising NP in
$M_{12}^{d,s}$) are not sufficient to absorb 
all discrepancies with the SM, namely the D\O\ measurement of 
$A_{\rm SL}$ and the  inconsistency between
$B(B\to\tau\nu)$ and $A_{\rm CP}^{\rm mix}(B_d\to J/\Psi K)$.
Still in Scenario I, which fits $\Delta_d$ and $\Delta_s$ independently,
we find the SM point $\Delta_d=\Delta_s=1$ disfavoured by  2.4$\,\sigma$;
this value was 3.6$\,\sigma$ in our 2010 analysis \cite{Lenz:2010gu} 
We notice that data still allow sizeable NP contributions  in both $B_d$ and $B_s$ sectors up to
30-40\% at the 3$\sigma$ level. 
The preference of Sc.~I over the SM mainly stems from the fact that 
$B(B\to\tau\nu)$ favours $\phi_d^\Delta<0$ which alleviates the problem
with $A_{\rm SL}$.  

In order to fully reconcile $A_{SL}$ with $\phi_{s}^{\psi \phi}$
we have extended our  study to a Scenario IV, which permits
  NP in both $M_{12}^{d,s}$ and $\Gamma_{12}^{d,s}$.  While this
scenario can accommodate all data, it is difficult to find realistic
models in which the preferred NP contributions to $\Gamma_{12}^s$
  (composed of Cabibbo-favoured tree-level decays) comply with other
  measurements. There are fewer phenomenological constraints on
  the Cabibbo-suppressed quantity $\Gamma_{12}^d$; a possible conflict
  with $M_{12}^d$ can be circumvented with chirality suppression.
  NP in $M_{12}^d$ and $\Gamma_{12}^d$ with the $B_s$ system
  essentially SM-like appears thus as an interesting possibility, requiring
  only a mild statistical upward fluctuation in the D\O\ data on $A_{\rm
    SL}$. Clearly,  independent measurements of $a^d_{\rm SL}$,
$a^s_{\rm SL}$ and/or $a^s_{\rm SL}-a^d_{\rm SL}$ are necessary
to determine whether scenarios with NP in $\Gamma_{12}^d$
  and/or $\Gamma_{12}^s$ are a viable explanation of
discrepancies in $\Delta F=2$ observables with respect to the Standard
Model.

\acknowledgments

We thank the CDF and LHCb collaborations for providing us with
  the 2D profile likelihood functions needed for our analyses. 
A.L.\ is supported by  DFG through a Heisenberg fellowship. U.N.\
acknowledges support by BMBF through grant 05H09VKF.

\end{document}